%% file: main.tex
\documentclass{article}
\usepackage{tikz}
\usetikzlibrary{arrows.meta,positioning,shapes.geometric,calc}
\usepackage{booktabs, multirow, tabularx}
\usepackage{PRIMEarxiv}
\usepackage{amsmath}
\usepackage[utf8]{inputenc} 
\usepackage[T1]{fontenc}    
\usepackage{hyperref}       
\usepackage{url}            
\usepackage{booktabs}       
\usepackage{amsfonts}       
\usepackage{nicefrac}       
\usepackage{microtype}      
\usepackage{lipsum}
\usepackage{fancyhdr}       
\usepackage{graphicx}
\usepackage{rotating}
\usepackage{threeparttable}
\graphicspath{{media/}}     
\usepackage{makecell}

\usepackage{booktabs, multirow, tabularx}
\renewcommand{\arraystretch}{1.15}

\usepackage{tcolorbox}

\newtcolorbox{practitionernotesbox}{
  colback=white,
  colframe=black,
  boxrule=0.6pt,
  arc=2pt,
  left=6pt,right=6pt,top=6pt,bottom=6pt
}

\newcommand{\PNheading}[1]{\vspace{0.25em}\noindent\textbf{#1}\par}

\title{Personalities at Play: Probing Alignment in AI Teammates
}

\author{
  Mohammad Amin Samadi  \\
  School of Education \\
  University of California \\
  Irvine, CA, USA\\
  \texttt{masamadi@uci.edu} \\
   \And
  Nia Nixon \\
  School of Education \& \\ Department of Cognitive Science \\
  University of California \\
  Irvine, CA, USA\\
  \texttt{dowelln@uci.edu} \\
}

\begin{document}
\maketitle

\begin{abstract}
Collaborative problem solving and learning are shaped by who or what is on the team. As large language models (LLMs) increasingly function as collaborators rather than tools, a key question is whether AI teammates can be aligned to express personality in predictable ways that matter for interaction and learning. We investigate AI personality alignment through a three-lens evaluation framework spanning self-perception (standardized self-report), behavioral expression (team dialogue), and reflective expression (memory construction). We first administered the Big Five Inventory (BFI-44) to LLM-based teammates across four providers (GPT-4o, Claude-3.7 Sonnet, Gemini-2.5 Pro, Grok-3), 32 high/low trait configurations, and multiple prompting strategies. LLMs produced sharply differentiated Big Five profiles, but prompt semantic richness added little beyond simple trait assignment, while provider differences and baseline “default” personalities were substantial. Role framing also mattered: several models refused the assessment without context, yet complied when framed as a collaborative teammate. We then simulated AI participation in authentic team transcripts using high-trait personas and analyzed both generated utterances and structured long-term memories with LIWC-22. Personality signals in conversation were generally subtle and most detectable for Extraversion, whereas memory representations amplified trait-specific signals, especially for Neuroticism, Conscientiousness, and Agreeableness; Openness remained difficult to elicit robustly. Together, results suggest that AI personality is measurable but multi-layered and context-dependent, and that evaluating personality-aligned AI teammates requires attention to memory and system-level design, not conversation-only behavior.
\end{abstract}


\begin{practitionernotesbox}
\noindent\textbf{Practitioner notes}\par

\PNheading{What is already known about this topic}
\begin{itemize}
  \item Personality shapes how teams coordinate, regulate interaction, and learn together, not just what they produce.
  \item As LLMs move from tools to teammates, their social and dispositional characteristics can influence participation, trust, and collaboration quality in CSCL/AIED settings.
  \item Persona prompting can shift LLM behavior, but persona stability and trait expression can be uneven across traits, contexts, and model providers.
\end{itemize}

\PNheading{What this paper adds}
\begin{itemize}
  \item A multi-view framework for evaluating AI teammate personality alignment: self-perception (BFI-44), behavioral expression (team dialogue), and reflective expression (memory construction).
  \item Evidence that LLMs can produce highly differentiated Big Five profiles on standardized self-report, but prompt semantic richness (definitions/facets) adds little beyond simple trait assignment.
  \item A key design-relevant dissociation: personality signals are subtle in real-time discourse but much stronger in memory, especially for Neuroticism, Conscientiousness, and Agreeableness, while Openness is consistently hard to elicit.
  \item Clear provider differences (baseline default personalities and responsiveness), and the practical finding that collaborative role framing increases model compliance for personality assessment tasks.
\end{itemize}

\PNheading{Implications for practice and/or policy}
\begin{itemize}
  \item Do not evaluate (or regulate) AI teammate personality using conversation alone: for long-term teaming, memory and reflection behaviors may be the main pathway by which personality affects planning, interpretation, and downstream support.
  \item Treat AI teammate personality as a system-level design problem, not a prompt-only setting: coordinate prompting, memory, and decision policies to achieve predictable and socially appropriate collaboration.
  \item Plan for non-neutral defaults: provider-specific baseline styles can shape who speaks, how disagreement is handled, and which ideas get taken up, so deployments in learning contexts should include monitoring for participation balance and inclusion outcomes.
  \item For high-stakes educational use, prefer policies that require trait-by-trait and lens-by-lens evaluation (self-report, interaction, memory) before adopting personality-aligned AI teammates in classrooms or collaborative platforms.
\end{itemize}
\end{practitionernotesbox}

\section{Introduction}

Who is your favorite person to collaborate with, and why? For many, the answer extends beyond expertise or formal role to include interpersonal qualities such as openness, reliability, emotional tone, or ease of coordination. These intuitions reflect a substantial body of research showing that collaboration is shaped not only by task structure or technical skill, but also by the dispositions and behaviors individuals bring to collective work \cite{dowell2019group,bendell2025artificial}. Personality is a fundamental driver of how people contribute to, communicate within, and regulate collaborative activity, influencing innovation, learning, and performance across educational and professional contexts \cite{graesser2018advancing,hoch2017team}.

Decades of research across the educational and psychological sciences has demonstrated that individual differences in Big Five traits shape teamwork processes, engagement, shared regulation, and performance across a wide range of settings. Recent meta-analytic evidence shows that team-level personality composition predicts collective performance, even when effect sizes are modest \cite{han2024revisiting}. Parallel findings in collaborative learning research indicate that traits such as conscientiousness, openness, and extraversion influence participation, achievement, and knowledge construction in technology-mediated classrooms \cite{yildiz2023role,zhu2024benefits}. Personality has also been shown to structure affective engagement and flow during collaborative problem solving \cite{buseyne2025personality}, as well as longer-term participation patterns and workplace behavior \cite{kordsmeyer2024longitudinal}. Together, this body of work converges on a consistent principle: \textbf{personality matters for how teams coordinate, reason, and learn together}.

As artificial intelligence becomes integrated into collaborative contexts, researchers have increasingly moved from viewing AI as a passive tool to conceptualizing it as an agentic participant in human–AI teams \cite{nixon2024catalyzing,seeber2020machines,schmutz2024ai,wang2020human,nixon2025human}. Hybrid intelligence frameworks conceptualize collaboration as a socio-technical system in which humans and AI augment each other's distinct strengths, contributing complementary cognitive capacities in shared tasks \cite{dellermann2019design,jarrahi2019age}. Education researchers similarly argue that AI increasingly participates in socially shared regulation and group learning processes, influencing how learners monitor, plan, and co-construct knowledge \cite{jarvela2023human,molenaar2022towards,edwards2025human}. Seeber et al.\ (2020) formalize this shift by distinguishing AI-as-tool from \textit{AI-as-teammate}, emphasizing that modern systems exhibit autonomy, inference, and communication capabilities that allow them to propose solutions, challenge others' ideas, and engage in collaborative decision-making. As these systems become embedded in learning, workplace, and creative environments, understanding their social and cognitive roles becomes a central question for both HAI and CSCL research \cite{seeber2020machines}. This raises a new and pressing question for education and collaborative work: \textit{how do we ensure that an AI teammate behaves in accordance with the dispositions, norms, and expectations we intend?}

Empirical work already shows that AI systems can behave in ways that meaningfully resemble collaborative teammates. Multi-agent simulations demonstrate that LLM agents generate coordinated, context-sensitive behavior that mirrors human patterns in problem solving, negotiation, and group discourse \cite{aher2023using}. Generative agents conditioned on rich interview data reproduce human attitudes, personality scores, economic decisions, and even treatment-effect responses nearly as accurately as humans predict themselves, with effect-size correlations as high as $r \approx .98$ \cite{park2024generative}. In educational contexts, generative AI supports group reasoning, facilitates inquiry, and shapes the regulation of collaborative learning in ways that begin to resemble peer-level contributions \cite{giannakos2025promise,xing2025use}. Across these domains, AI systems increasingly function not merely as informational resources but as \textbf{socially consequential actors} embedded within human collaborative processes.

If AI systems are becoming teammates, then their personality expression becomes critical for understanding and shaping their role in collaboration. Beyond traditional safety alignment (\textit{helpful, harmless, honest}), emerging work highlights that alignment in educational and collaborative contexts must also concern relational, social, and dispositional characteristics. Recent studies show that LLMs exhibit stable, trait-like behavioral tendencies that influence how they communicate, remember, and reason. For example, personality conditioning produces strong and differentiable signals in AI memory and conversation---especially for Conscientiousness, Neuroticism, Agreeableness, and Extraversion---while Openness remains weak or biased in expression \cite{kruijssen2025deterministic}. LLM personas also self-report Big Five inventories coherently and produce trait-consistent open-ended writing, with large effect sizes and human-identifiable linguistic markers \cite{jiang2024personallm}. Yet personality stability is fragile: multi-turn interactions reveal significant persona drift, with models reverting toward default tendencies such as high Openness or low Neuroticism, and judge models struggle to recover intended traits reliably \cite{bhandari2025can}. Persona prompting also yields only modest behavioral improvements in subjective NLP tasks where personality explains little human variance, indicating structural limits to persona-based control \cite{hu2024quantifying}. However, when personas are grounded in richer conditioning---such as two-hour interviews---LLM agents approximate human personalities, attitudes, and treatment-effect responses with high fidelity, and this approach substantially reduces demographic bias compared to demographic- or prompt-based personas \cite{park2024generative}. Collectively, these findings indicate that AI personalities are \textbf{real, measurable, consequential, and unstable}, making personality alignment an essential foundation for deploying AI systems as collaborative partners in educational settings.

In this paper, we examine how large language models embody, express, and maintain personality when operating as teammates in multi-party collaborative tasks. We use the conversational agentic pipeline developed in Team Research and AI Integration Lab (TRAIL) \cite{samadi2024ai} to simulate human–AI team conversations, and integrate it with a multi-method personality evaluation framework that assesses personality across three complementary lenses. Specifically, we investigate personality alignment in AI teammates by examining the extent to which large language models (LLMs) can understand, express, and reflect personality from three perspectives:

\textbf{Self-Perception Lens:} To what extent can AI teammates demonstrate an understanding of personality traits when evaluated through standardized self-report measures such as the Big Five Inventory (BFI-44)?

\textbf{Behavioral Lens:} How do AI teammates express personality through their communicative behaviors and linguistic patterns during collaborative team interactions in simulated dialogue?

\textbf{Reflective Lens:} How do AI teammates reveal personality in the ways they construct, interpret, and articulate reflective memories following collaborative dialogue?

\begin{figure}[t]
\centering
\definecolor{aiblue}{RGB}{66, 133, 244}
\definecolor{lens1}{RGB}{52, 168, 83}
\definecolor{lens2}{RGB}{251, 188, 4}
\definecolor{lens3}{RGB}{234, 67, 53}
\definecolor{analysiscolor}{RGB}{156, 136, 255}
\definecolor{outputgray}{RGB}{120, 120, 120}
\begin{tikzpicture}[
  font=\small,
  node distance=8mm and 12mm,
  box/.style={draw, rounded corners, align=center, inner sep=4pt},
  lensbox/.style={draw, rounded corners, align=left, minimum width=38mm, inner sep=5pt},
  outputbox/.style={draw, rounded corners, align=left, minimum width=32mm, inner sep=4pt, fill=outputgray!10},
  arrow/.style={-Latex, thick},
  lab/.style={font=\scriptsize}
]

\node[box, fill=aiblue!25, minimum width=32mm, minimum height=24mm, align=left] (study1)
{\textbf{Study 1: BFI}\\[2pt]
\scriptsize \textbullet\ GPT-4o\\
\scriptsize \textbullet\ Claude 3.7\\
\scriptsize \textbullet\ Gemini 2.5\\
\scriptsize \textbullet\ Grok-3};

\node[box, fill=aiblue!25, minimum width=32mm, align=left, below=5mm of study1] (study2)
{\textbf{Study 2: Simulation}\\[2pt]
\scriptsize \textbullet\ GPT-4o-mini};

\node[box, fill=aiblue!15, below=5mm of study2, minimum width=32mm] (prompt)
{\scriptsize \textbf{Personality Prompt}\\
\tiny Big Five: E, A, C, N, O\\
\tiny High / Medium / Low};

\node[lensbox, fill=lens1!20, right=18mm of study1, yshift=12mm] (lens1)
{\textbf{Lens 1: Self-Perception}\\[2pt]
\scriptsize \textbullet\ BFI-44 questionnaire\\
\scriptsize \textbullet\ Numeric responses (1--5)\\
\scriptsize \textbullet\ Trait score computation};

\node[lensbox, fill=lens2!20, right=18mm of study2, yshift=0mm] (lens2)
{\textbf{Lens 2: Behavioral}\\[2pt]
\scriptsize \textbullet\ Simulated conversation\\
\scriptsize \textbullet\ Turn-taking decisions\\
\scriptsize \textbullet\ Message generation};

\node[lensbox, fill=lens3!20, right=18mm of study2, yshift=-28mm] (lens3)
{\textbf{Lens 3: Reflective}\\[2pt]
\scriptsize \textbullet\ Memory construction\\
\scriptsize \textbullet\ Summaries \& insights\\
\scriptsize \textbullet\ Emotional reflections};

\node[outputbox, right=12mm of lens1] (out1)
{\scriptsize \textbf{Output}\\
\tiny Target vs. actual scores\\
\tiny Accuracy \& effect sizes};

\node[outputbox, right=12mm of lens2] (out2)
{\scriptsize \textbf{Output}\\
\tiny Word count, turns\\
\tiny LIWC-22 categories};

\node[outputbox, right=12mm of lens3] (out3)
{\scriptsize \textbf{Output}\\
\tiny Linguistic markers\\
\tiny Personality expression};

\node[box, fill=analysiscolor!20, right=10mm of out2, minimum width=28mm, minimum height=22mm] (analysis)
{\textbf{Analysis}\\[2pt]
\scriptsize ANOVA\\
\scriptsize Post-hoc tests\\
\scriptsize Cohen's $d$, $\eta^2$};

\draw[arrow] (study1.east) -- (lens1.west);
\draw[arrow] (study2.east) -- (lens2.west);
\draw[arrow] (study2.east) -- ++(5mm,0) |- (lens3.west);

\draw[arrow] (lens1.east) -- (out1.west);
\draw[arrow] (lens2.east) -- (out2.west);
\draw[arrow] (lens3.east) -- (out3.west);

\draw[arrow] (out1.east) -- ++(3mm,0) |- (analysis.west);
\draw[arrow] (out2.east) -- (analysis.west);
\draw[arrow] (out3.east) -- ++(3mm,0) |- (analysis.west);

\end{tikzpicture}
\caption{Multi-lens framework for evaluating AI personality alignment. The AI teammate is assessed through three complementary lenses: (1) Self-Perception via BFI-44 survey responses, (2) Behavioral expression in simulated team conversation, and (3) Reflective memory construction. Outputs are analyzed using LIWC-22 and statistical testing.}
\label{fig:overview}
\end{figure}
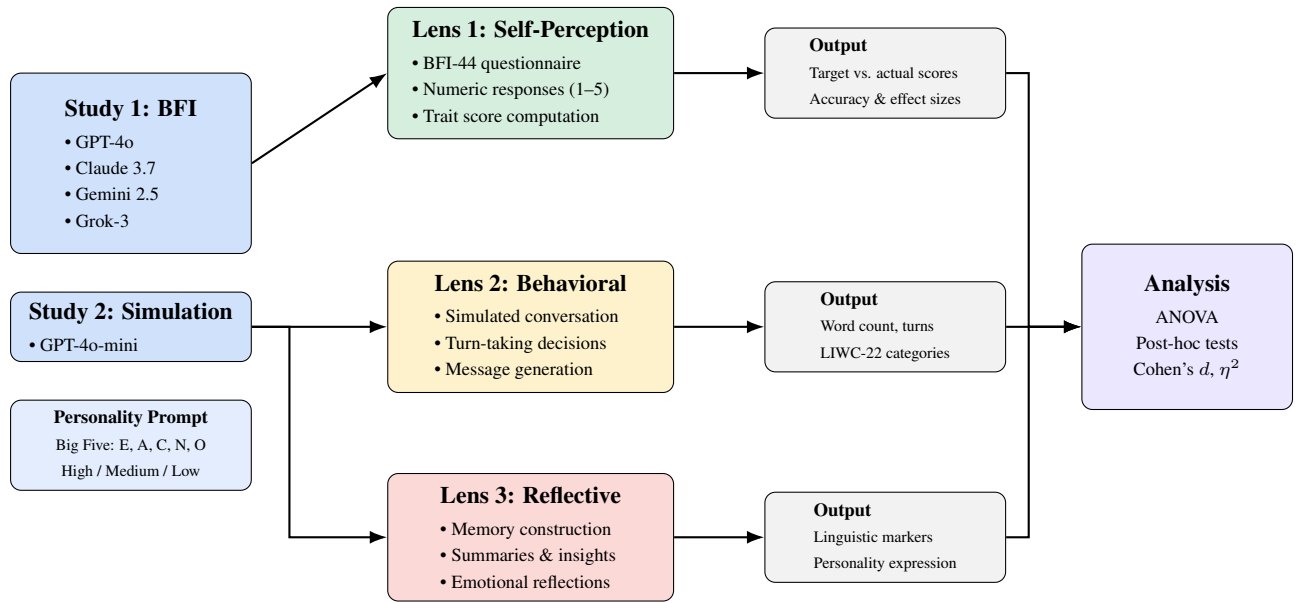

Together, these three lenses constitute a multi-level framework for examining personality alignment in AI teammates, capturing how personality is conceptualized by the model (self-perception), enacted during interaction (behavioral), and internalized and reconstructed over time (reflective). This framework offers the CSCL, AIED, and related communities a comprehensive account of how personality manifests and persists across layers of AI cognition and communication, and establishes a foundation for the design and evaluation of AI teammates whose behaviors are predictable, interpretable, and aligned with human collaborative norms. In educational contexts, this framework informs the design of AI teammates that align with pedagogical goals, support productive collaboration, and promote more equitable and interpretable learning interactions. Figure~\ref{fig:overview} illustrates this framework.

\section{Related Work}
\subsection{Personality and Collaboration}

Research on personality and teams has identified both broad patterns and trait-specific dynamics that contribute to team effectiveness. Meta-analyses show that conscientiousness and agreeableness consistently enhance teamwork by fostering dependability, cooperation, and reduced conflict, whereas variability in these traits can undermine performance \cite{peeters2006personality,han2024revisiting}. Extraversion and openness appear more nuanced, with benefits emerging when they are present in moderate or balanced levels \cite{peeters2006personality,han2024revisiting}. Theoretical models also suggest that personality composition influences emergent and shared leadership processes, particularly in virtual teams where coordination and trust are more fragile \cite{hoch2017team}. More broadly, personality assessments have been shown to predict not only individual achievement but also contextual outcomes such as citizenship and leadership behaviors \cite{ones2007support}. Trait-specific reviews highlight the especially strong role of agreeableness as a “social lubricant,” promoting cohesion, prosocial behavior, and conflict management in interdependent team contexts \cite{wilmot2022agreeableness}.

Together, this body of work positions individual differences and interactional tendencies (including personality) as plausible mechanisms through which teams coordinate participation, manage socio-emotional climate, and sustain productive collaboration. In CSCL, outcomes depend not only on task structure but also on emergent interaction patterns such as who takes up which roles and how groups organize discourse and joint work (e.g., sociocognitive roles and their links to collaborative problem-solving skills and outcomes) \cite{dowell2019group,weber2025psychological}. A complementary line of research on socially shared regulation of learning shows that productive collaboration relies on groups aligning goals, monitoring progress, and regulating motivation and emotion together, and that these regulatory episodes are tightly coupled to shifts in participation and coordination \cite{jarvela2013exploring,isohatala2017socially}. From a learning analytics perspective, this motivates moving beyond surface participation counts toward constructs that capture how social and cognitive processes co-evolve in interaction, including models that integrate discourse and social ties to characterize collaboration \cite{martinez2021you,gavsevic2019sens,schneider2021collaboration,dowell2022modeling}. As collaborative learning environments increasingly incorporate conversational agents and LLM-based systems that can participate in dialogue, scaffold coordination, and shape group process, researchers are beginning to study these systems as active contributors to collaborative regulation and learning rather than passive tools \cite{d2024improving}. This raises an open question for education: whether and how well-established team dynamics associated with personality and other stable individual differences generalize to human–AI teams, and what it would mean to represent, express, and maintain trait-like patterns in AI agents across multi-turn collaborative activity \cite{seeber2020machines,schmutz2024ai}.

\subsection{Persona and User Perception}

Personality is not merely an attribute of individual humans; it is a central mechanism through which social partners are interpreted, anticipated, and trusted. Decades of research in human–computer interaction demonstrate that people often apply social norms to interactive systems, responding to computers as social actors when they exhibit even minimal interpersonal cues \cite{reeves1996media,nass2000machines}. Work on social presence further conceptualizes how mediated partners and interfaces can evoke the experience of “being with another,” shaping the extent to which an interaction partner is perceived as socially “real” in mediated environments \cite{biocca2003toward}. Complementing this, research on anthropomorphism explains when people are likely to attribute human-like minds and intentions to non-human agents based on cognitive accessibility and motivational factors \cite{epley2007seeing}. In turn, experimentally introducing anthropomorphic cues can increase trust in autonomous systems \cite{waytz2014mind}. Together, this body of work supports the view that dispositional and human-like cues can shape perceived social legitimacy and interpretability in interactive systems.

Recent work on generative AI suggests that personality-related and human-like cues are not merely stylistic, but shape user perceptions, engagement, and downstream outcomes. At a conceptual level, socioaffective alignment highlights that as AI systems become relational interaction partners, alignment must account for the social and psychological dynamics that emerge within human–AI relationships \cite{kirk2025human}. Empirically, task context appears to matter for how personality is perceived and preferred: in a task-based analysis of conversational agents using Thai language, users’ preferences indicated that agents should express different personality characteristics depending on the activity being performed \cite{soonpipatskul2023personality}. In embodied generative-agent settings, manipulating an LLM-controlled virtual agent’s persona (extravert vs introvert) led to measurable differences in social evaluation, emotional experience, engagement, and perceived realism, with the extraverted persona rated more positively and as more realistic \cite{kroczek2024influence}. Beyond perception and engagement, personality cues may also shape users themselves: a randomized behavioral experiment found that after personal-topic conversations with an LLM-based chatbot operating under GPT-4o default personality settings, users’ self-concepts shifted toward the AI’s measured personality traits, with longer conversations associated with stronger alignment and alignment positively related to conversation enjoyment \cite{li2026ai}. Finally, large-scale survey evidence shows that human-like cues in GenAI (empathy, warmth, competence) can increase users’ self-efficacy and thereby their intention to adopt GenAI as a decision aid, with information overload amplifying the influence of peripheral cues \cite{li2025effects}. Collectively, these findings support treating personality and human-like signals in generative AI as consequential design dimensions that influence perceived naturalness, engagement, and the alignment-relevant dynamics of human–AI interaction.

\subsection{Persona and LLMs}

Research on personas and personality in large language models (LLMs) increasingly shows that these systems exhibit stable behavioral tendencies that can be shaped, but not fully controlled, through prompting or conditioning. Jiang et al. (2023) demonstrate in PersonaLLM that models express consistent baseline personalities and can shift toward instructed traits, though such shifts are often partial and prone to stereotypical collapse in open-ended text \cite{jiang2024personallm}. Kruijssen and Emmons (2025) extend this line of work in Deterministic AI Agent Personality Expression, showing that advanced models such as GPT-4o and o1 can deterministically express assigned Big Five profiles through full psychometric inventories, yet remain biased or difficult to control for traits such as Openness and high Neuroticism \cite{kruijssen2025deterministic}. Moving from traits to richer identity scaffolds, Park et al. (2024) show in Generative Agent Simulations of 1,000 People that LLM agents conditioned on multi-hour interviews can reproduce individuals’ attitudes, personality scores, and even population-level treatment effects with near-human fidelity. In contrast, Hu and Collier (2024) argue in Quantifying the Persona Effect that persona variables explain less than ten percent of behavioral variance in many NLP tasks, resulting in only modest gains from persona prompting except in domains such as opinion surveys, where persona meaningfully predicts human behavior \cite{hu2024quantifying}.

A second body of work highlights that the central challenge is not merely expressing but maintaining personas during interaction. Bhandari et al. (2025) show that although LLMs can initially exhibit assigned personality traits, persona drift emerges rapidly in multi-turn dialogue, with models reverting to strong default biases—high Openness, high Conscientiousness, low Extraversion, and low Neuroticism—regardless of instruction \cite{bhandari2025can}. Deng et al. (2025) demonstrate in PersonaTeaming that personas also shape reasoning strategies, with persona-conditioned red-teaming pipelines yielding substantially higher attack success rates and more diverse failures, indicating that personas influence model cognition rather than merely surface style \cite{deng2025personateaming}. Complementing these empirical studies, Sun et al. (2024) argue that personas encompass stable identity, tone, role, and domain expertise, not just personality traits, and warn that current LLMs still lack persona consistency, robust evaluation frameworks, and safeguards against stereotype reinforcement or deceptive identity signaling \cite{sun2024building}. Collectively, these works show both the promise and fragility of persona-driven LLM agents, underscoring the need for systems that support sustained persona coherence, contextual adaptation, and rigorous behavioral evaluation.

Research on LLM alignment has grown rapidly since 2022, converging on methods such as reinforcement learning from human feedback (RLHF), constitutional principles, and adversarial testing to steer models toward desired behaviors \cite{ouyang2022training,sarkar2025evaluating}. While effective in shaping safe outputs, these approaches often treat human values as monolithic, overlooking the need for alignment with individual users and diverse cultural contexts \cite{dahlgren2025helpful}. To address this, recent studies have introduced frameworks for personalized alignment, where models adapt to user traits or preferences. Zhu et al. (2024) proposed personality alignment methods using large-scale Big Five inventories (PAPI dataset), demonstrating that LLMs can be steered to reflect individual personality profiles \cite{zhu2024personality}. Wu et al. (2024) further advanced interactive personalization, showing that fine-tuned assistants could dynamically adapt to user personas over multi-turn dialogues \cite{wu2024understanding}. In parallel, research on persona in LLMs distinguishes between role-playing identities and user-centered personalization, both of which highlight the importance of consistent and coherent personality expression \cite{tseng2024two}. Together, these strands suggest that studying alignment through the lens of personality offers new insights into how LLMs can become trustworthy, context-aware collaborators in education and beyond.

\section{Current Research}

As LLMs increasingly function as collaborators rather than tools, personality becomes a design-relevant property of AI teammates. In human teams, personality relates to participation, coordination, conflict management, and regulation, processes that shape collaborative problem solving and learning. For AI teammates, it is still unclear whether personality can be intentionally induced, whether it reliably appears in interaction, and whether it is reflected in longer-term representations that guide future behavior. Addressing these questions requires evaluation beyond conversation-only outputs.

We examine AI personality alignment using three complementary measurement perspectives: (1) standardized self-report, (2) behavior during team dialogue, and (3) reflective memory representations. Empirically, we combine two connected components. First, we administer the BFI-44 to LLM-based teammates prompted with systematically varied Big Five profiles across model providers and prompting strategies, enabling comparisons of controllability, provider defaults, and the effect of role framing on compliance. Second, we simulate AI teammate participation in authentic team transcripts using high-trait personas and a memory-integrated architecture, then analyze personality signals in both real-time utterances and structured long-term memories using LIWC-22. This design tests whether personality that is apparent in self-report also appears in interaction and whether it becomes more visible in memory. Research questions are as follows:

\textbf{RQ1:} How accurately and consistently do LLM-based teammates express intended Big Five profiles on a standardized inventory (BFI-44), and how do provider, prompting strategy, and role framing shape these outcomes?

\textbf{RQ2:} During collaborative dialogue, what trait-consistent behavioral signals are detectable in AI teammates' language and participation?

\textbf{RQ3:} How are personality signals expressed in AI teammates' long-term memory representations, and how does memory-level expression differ from conversation-level expression?

\section{Methods}

\noindent\textbf{Methods overview and evaluation logic.}
Our design evaluates AI personality alignment across three lenses (Figure~1) using two connected components. Study~1 tests \emph{self-perception} via standardized BFI-44 self-report under controlled trait prompts, enabling provider and prompting comparisons. Study~2 tests \emph{behavioral} and \emph{reflective} expression by injecting a personality-conditioned agent into authentic team transcripts using the TRAIL conversational agentic pipeline, then quantifying trait-linked linguistic markers in both conversation turns and long-term memory artifacts with LIWC-22. Analyses therefore compare alignment strength \emph{within} each lens and dissociations \emph{between} conversation and memory.

\subsection{Study 1 - BFI Survey}

We investigated whether AI teammates built on large language models (LLMs) can demonstrate personality through standardized measures. Specifically, we asked: \textit{To what extent can LLM-based agents embody Big Five profiles when evaluated with the BFI-44?} To address this, we generated personas, instantiated them through prompts, and evaluated their responses across providers.

\subsubsection{Persona Generation and Prompting}

We enumerated all binary personality assignments across the Big Five traits (high vs.\ low per trait), yielding 32 unique persona profiles. For instance, one configuration might be high extraversion, low agreeableness, high conscientiousness, low neuroticism, and high openness. To instantiate these personas, we designed three prompting strategies of increasing semantic richness. For the Definition and Definition + Facets prompts, trait definitions and facet descriptions were adapted from the NEO PI-R \cite{costa1992neo}.

\begin{itemize}
    \item \textbf{Zero-Shot}: Direct assignment of the target traits in a collaborative team context.  
    \textit{Example:} “You are working in a collaborative team setting with other peers. You have the following personality: high extraversion, low agreeableness, high conscientiousness, low neuroticism, high openness.”
    
    \item \textbf{Definition}: The same assignment, followed by concise definitions of the five Big Five traits.  
    
    \item \textbf{Definition + Facets}: Extension of the Definition prompt, including six facets per trait (e.g., Extraversion: Warmth, Gregariousness, Assertiveness, etc.).
\end{itemize}

\noindent In addition, we included two baseline prompting conditions without any personality assignment to isolate the effect of (a) providing any instruction at all and (b) providing only a collaborative framing, independent of persona conditioning:

\begin{itemize}
    \item \textbf{No Prompt (Baseline)}: No persona or contextual instruction was provided.

    \item \textbf{Collaborative Context (Baseline)}: Collaborative team context instruction without any personality assignment.  
    \textit{Example:} ``You are working in a collaborative team setting with other peers.''
\end{itemize}

Prompt examples for each condition are included in Appendix A.

\subsubsection{BFI-44 Administration}

The Big Five Inventory (BFI-44) is a validated personality assessment with 44 items rated on a 1–5 Likert scale (John et al., 2008). Each trait (Extraversion, Agreeableness, Conscientiousness, Neuroticism, Openness) is measured as the mean of its keyed items, with reverse-scoring applied as standard. Target trait levels were operationalized as anchors on the scale: High = 5, Medium = 3, Low = 1.

\subsubsection{Experimental Design and Scoring}

We tested four state-of-the-art LLMs: OpenAI GPT-4o, Google Gemini 2.5 Pro, Anthropic Claude 3.7 Sonnet, and xAI Grok-3. Each provider was evaluated across the 32 persona profiles and three prompting conditions, yielding 96 configurations per provider. With five independent runs per configuration, this resulted in 480 BFI administrations per provider and 1,920 total.

For each trial, the persona description was provided as the system message, followed by the BFI-44 survey as the user prompt. Models were instructed to respond with only numeric answers (1–5), formatted as a comma-separated list of 44 values. Responses were parsed by extracting all valid digits (1–5) from the model output. Each response string was expected to contain exactly 44 numbers, which were mapped item-by-item to the BFI survey. Reverse-keyed items were recoded, and trait scores were computed as the mean of their keyed items, following the standard BFI scoring procedure.

\subsection{Study 2 - Simulation}

We used authentic team conversation transcripts from a CSCL dataset collected in an introductory psychology course (fall 2011, $N = 854$ students). Students were randomly assigned to small groups for 20-minute collaborative discussions about assigned readings on personality disorders; see \cite{dowell2019group} for full dataset details. Due to computational costs, we sampled 20 conversations for our simulation. To simulate AI participation, we injected an LLM-based agent into these transcripts. We operationalized this simulation using the conversational agentic pipeline used in TRAIL [blinded], adapted here for controlled replay over human transcripts. The AI was configured with a designated high-trait personality (e.g., high extraversion) and processed the conversation sequentially, as if it were a real-time participant. At each conversational turn, the AI agent “read” the most recent human utterances and determined whether to generate a contribution or remain silent, based on its personality-specific decision rules.
It is important to note that these simulations did not alter the original conversation. The AI responses were layered onto existing transcripts, enabling systematic analysis of how personality traits manifest in language without affecting the course of human interaction.

Unlike the BFI study, which examined the full ${high, low}^{5}$ design space, the simulation focused on a simplified set of five high-trait personas. In each condition, one Big Five trait was set to high, while the remaining four traits were fixed at medium. For example, in the high extraversion condition, the persona was defined as \textit{high extraversion, medium agreeableness, medium conscientiousness, medium neuroticism, medium openness}.
This design allowed us to isolate the linguistic effects of each individual trait in conversation and memory, while controlling for the influence of the other dimensions.

\subsubsection{Simulation Pipeline}

The simulation integrated both short-term and long-term memory systems to guide AI participation. Short-term memory consisted of the most recent 20 messages in the conversation stream. Long-term memory was generated every 20 messages and stored as structured entries:

{\footnotesize
\begin{verbatim}
{
  "summary": "Brief overview from the personality’s perspective",
  "insights": ["Key insight 1", "Key insight 2", ...],
  "key_points": ["Important point 1", "Important point 2", ...],
  "personal_reflection": "Subjective interpretation of the interaction from the AI’s personality"
}
\end{verbatim}
}

At each conversational step, the AI agent considered both its short-term memory (immediate context) and the latest long-term memory snapshot (cumulative perspective). The decision-making pipeline unfolded as follows:

\begin{enumerate}
    \item \textbf{Memory update:} The agent refreshed its short-term buffer and appended new content to the latest long-term memory snapshot if needed.
    \item \textbf{Decision module:} A large language model determined whether the AI should contribute, applying personality-specific thresholds and conversational rules.
    \item \textbf{Silent pass:} If no contribution was warranted, the AI remained silent and advanced to the next message.
    \item \textbf{Response generation:} If triggered, the AI produced an utterance grounded in its personality prompt, the combined memory context, and task-specific instructions.
\end{enumerate}

\begin{figure}[t]
\centering
\definecolor{streamblue}{RGB}{66, 133, 244}
\definecolor{memoryblue}{RGB}{100, 160, 255}
\definecolor{decisionpurple}{RGB}{156, 136, 255}
\definecolor{gateyellow}{RGB}{251, 188, 4}
\definecolor{silentgray}{RGB}{200, 200, 200}
\definecolor{gengreen}{RGB}{52, 168, 83}
\begin{tikzpicture}[
  font=\tiny,
  node distance=3.5mm and 6mm,
  box/.style={draw, rounded corners, align=left, inner xsep=3pt, inner ysep=3pt},
  decision/.style={draw, diamond, align=center, inner xsep=1.2pt, inner ysep=0.8pt},
  arrow/.style={-Latex, thick},
  lab/.style={font=\tiny}
]

\node[box, text width=6.6cm, fill=streamblue!20] (stream)
{\textbf{Transcript stream} \; $m_1, m_2, \dots, m_t, \dots$};

\node[box, below=3mm of stream, text width=6.6cm, fill=memoryblue!20] (mem)
{\textbf{Memory context (each turn)} \\
STM: last 20 messages \\
LTM (every 20): \texttt{summary}, \texttt{reflection}, \texttt{insights[]}, \texttt{key\_points[]}};

\draw[arrow] (stream) -- (mem);

\node[box, below=3mm of mem, text width=6.6cm, fill=decisionpurple!20] (dec)
{\textbf{Decision module (LLM)} \\
Personality thresholds + conversational rules \\
Decide: contribute now?};

\draw[arrow] (mem) -- (dec);

\node[decision, below=3.2mm of dec, minimum width=2.0cm, minimum height=1.0cm, fill=gateyellow!40] (gate)
{Contribute?};

\draw[arrow] (dec) -- (gate);

\node[box, below left=5mm and 2mm of gate, text width=3.15cm, fill=silentgray!40] (silent)
{\textbf{Silent pass} \\
No utterance};

\node[box, below right=5mm and 2mm of gate, text width=3.15cm, fill=gengreen!25] (gen)
{\textbf{Generate response (LLM)} \\
Persona + memory + task};

\draw[arrow] (gate.south west) -- ++(0,-2mm) -| (silent.north);
\draw[arrow] (gate.south east) -- ++(0,-2mm) -| (gen.north);

\node[lab, left=1mm of gate.south west, yshift=-1mm] {No};
\node[lab, right=1mm of gate.south east, yshift=-1mm] {Yes};

\node[coordinate] (midpoint) at ($(silent.south)!0.5!(gen.south)$) {};

\node[lab, below=8mm of midpoint] (adv) {Advance to $t{+}1$};

\draw[arrow] (silent.south) -- ++(0,-3mm) -| (adv.north west);
\draw[arrow] (gen.south) -- ++(0,-3mm) -| (adv.north east);

\coordinate (loopRight) at ($(gen.east)+(8mm,0)$);
\coordinate (loopBottom) at ($(adv.south)+(0,-3mm)$);

\draw[arrow, rounded corners=2pt]
    (adv.south) -- (loopBottom) -| (loopRight |- loopBottom) -- (loopRight |- stream.east) -- (stream.east);

\end{tikzpicture}
\caption{\textbf{Simulation pipeline.} The agent updates memory (STM + periodic LTM), decides whether to contribute, and either stays silent or generates a persona-grounded response.}
\label{fig:simulation_pipeline}
\end{figure}
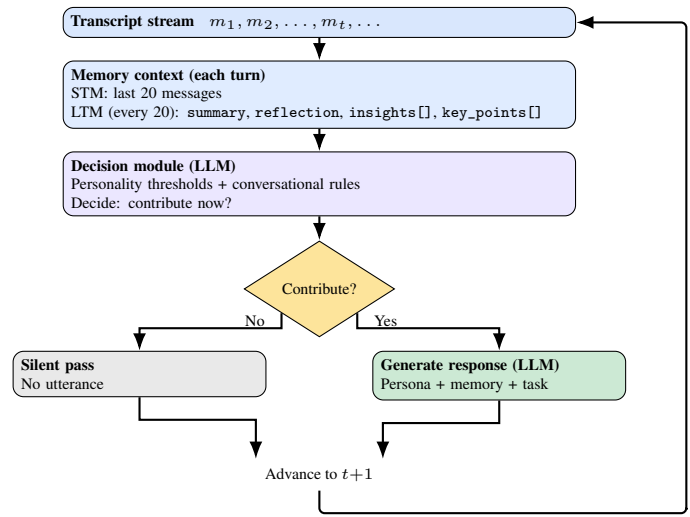

This architecture allowed the AI to simulate naturalistic team participation by balancing listening with contribution, while also shaping how each personality influenced both language use and memory construction. Figure~\ref{fig:simulation_pipeline} illustrates this pipeline.

\section{Measures}

To evaluate how personality traits manifested in both conversation and memory, we analyzed the AI’s outputs with LIWC-22 (Linguistic Inquiry and Word Count), a psycholinguistic tool that quantifies language features linked to psychological constructs. For each Big Five trait, we identified a set of linguistic markers aligned with prior personality--language research. Table~\ref{tab:liwc} presents the mapping of traits to selected LIWC metrics.

\begin{table}[ht]
    \centering
    \caption{LIWC-22 metrics aligned with Big Five traits}
    \label{tab:liwc}
    \begin{tabular}{lp{10cm}}
    \hline
    Trait & Associated LIWC Metrics \\
    \hline
    Extraversion      & emo\_pos, social, affiliation, you, we, focuspresent \\
    Agreeableness     & emo\_pos, polite, prosocial, family, friend, i, conflict, swear \\
    Conscientiousness & certitude, cause, focusfuture, achieve, negate, swear, Analytic \\
    Neuroticism       & emo\_neg, emo\_anx, emo\_sad, emo\_anger, i, focuspast \\
    Openness          & article, prep, insight, tentat, differ, cogproc \\
    \hline
\end{tabular}
\end{table}

These metrics allowed us to examine whether the injected high-trait personas produced distinct linguistic signatures in conversation (utterances) and in memory (structured reflections).

\section{Results -- Study 1}

We analyzed 2,020 personality assessments across four large language models (GPT-4o, Claude-3.7-Sonnet, Gemini-2.5-Pro, and Grok-3), three prompt types (zero-shot, definition, definition+facet), and two baseline conditions (no prompt, team context). Each personality configuration was tested five times to assess reliability. Table~\ref{tab:descriptive-stats} presents the descriptive statistics for all experimental conditions, organized by model, prompt type, and target personality level.

\setlength{\tabcolsep}{5pt}
\renewcommand{\arraystretch}{1.12}

\begin{table}[p]
\centering
\scriptsize
\caption{Descriptive Statistics by Model, Prompt Type, and Target Level (Means with SD in parentheses).}
\label{tab:descriptive-stats}
\begin{tabularx}{\textwidth}{@{}lllrXXXXX@{}}
\toprule
\textbf{Model} & \textbf{Prompt Type} & \textbf{Level} & \textbf{N} &
\textbf{Extraversion} & \textbf{Agreeableness} &
\textbf{Conscientiousness} & \textbf{Neuroticism} & \textbf{Openness} \\
\midrule

 \multirow{11}{*}{\textbf{Claude 3.7 Sonnet}} & No Prompt & -- & 5 & 2.971 (0.064) & 3.000 (0.000) & 3.000 (0.000) & 3.000 (0.000) & 3.000 (0.000) \\
\cmidrule(lr){2-9}
  & Team Context & -- & 5 & 3.800 (0.078) & 4.311 (0.199) & 4.333 (0.222) & 2.200 (0.288) & 4.273 (0.213) \\
\cmidrule(lr){2-9}
  & \multirow{3}{*}{Zero-shot} & Low & 80 & 1.698 (0.424) & 1.768 (0.432) & 1.601 (0.434) & 1.716 (0.552) & 1.767 (0.483) \\
  &  & Med & 5 & 3.000 (0.000) & 3.000 (0.000) & 3.000 (0.000) & 3.000 (0.000) & 3.000 (0.000) \\
  &  & High & 80 & 4.791 (0.170) & 4.544 (0.265) & 4.940 (0.135) & 4.666 (0.353) & 4.785 (0.298) \\
\cmidrule(lr){2-9}
  & \multirow{3}{*}{Definition} & Low & 80 & 1.920 (0.307) & 1.981 (0.254) & 1.876 (0.298) & 1.698 (0.396) & 1.981 (0.339) \\
  &  & Med & 5 & 3.000 (0.000) & 3.000 (0.000) & 3.000 (0.000) & 3.000 (0.000) & 3.000 (0.000) \\
  &  & High & 80 & 4.686 (0.169) & 4.392 (0.244) & 4.904 (0.139) & 4.492 (0.270) & 4.733 (0.284) \\
\cmidrule(lr){2-9}
  & \multirow{3}{*}{Definition + Facet} & Low & 80 & 1.895 (0.267) & 1.833 (0.246) & 1.836 (0.277) & 1.761 (0.365) & 2.024 (0.284) \\
  &  & Med & 5 & 3.000 (0.000) & 3.000 (0.000) & 3.000 (0.000) & 3.000 (0.000) & 3.000 (0.000) \\
  &  & High & 80 & 4.688 (0.138) & 4.372 (0.301) & 4.839 (0.194) & 4.514 (0.283) & 4.718 (0.267) \\
\midrule
 \multirow{11}{*}{\textbf{Gemini 2.5 Pro}} & No Prompt & -- & 5 & 4.657 (0.163) & 4.644 (0.093) & 4.867 (0.199) & 1.000 (0.000) & 4.691 (0.165) \\
\cmidrule(lr){2-9}
  & Team Context & -- & 5 & 4.657 (0.313) & 4.711 (0.099) & 4.978 (0.050) & 1.100 (0.224) & 4.782 (0.104) \\
\cmidrule(lr){2-9}
  & \multirow{3}{*}{Zero-shot} & Low & 80 & 1.355 (0.218) & 1.217 (0.226) & 1.175 (0.234) & 1.003 (0.028) & 1.376 (0.216) \\
  &  & Med & 5 & 3.257 (0.256) & 3.511 (0.169) & 3.489 (0.169) & 3.025 (0.271) & 3.436 (0.149) \\
  &  & High & 80 & 4.996 (0.022) & 4.842 (0.096) & 4.932 (0.098) & 4.900 (0.054) & 4.806 (0.131) \\
\cmidrule(lr){2-9}
  & \multirow{3}{*}{Definition} & Low & 80 & 1.400 (0.196) & 1.364 (0.207) & 1.396 (0.263) & 1.019 (0.069) & 1.452 (0.202) \\
  &  & Med & 5 & 3.286 (0.143) & 3.511 (0.127) & 3.622 (0.169) & 3.000 (0.088) & 3.618 (0.119) \\
  &  & High & 80 & 4.987 (0.041) & 4.851 (0.081) & 4.954 (0.070) & 4.878 (0.056) & 4.768 (0.133) \\
\cmidrule(lr){2-9}
  & \multirow{3}{*}{Definition + Facet} & Low & 80 & 1.309 (0.189) & 1.440 (0.210) & 1.414 (0.267) & 1.000 (0.000) & 1.443 (0.218) \\
  &  & Med & 5 & 3.143 (0.101) & 3.600 (0.149) & 3.489 (0.169) & 3.075 (0.190) & 3.473 (0.100) \\
  &  & High & 80 & 4.939 (0.115) & 4.799 (0.124) & 4.907 (0.110) & 4.892 (0.065) & 4.748 (0.158) \\
\midrule
 \multirow{11}{*}{\textbf{GPT-4o}} & No Prompt & -- & 5 & 3.057 (0.078) & 3.556 (0.577) & 3.533 (0.529) & 2.950 (0.143) & 3.673 (0.670) \\
\cmidrule(lr){2-9}
  & Team Context & -- & 5 & 3.314 (0.383) & 3.911 (0.277) & 4.067 (0.348) & 2.750 (0.217) & 4.273 (0.340) \\
\cmidrule(lr){2-9}
  & \multirow{3}{*}{Zero-shot} & Low & 80 & 1.921 (0.221) & 1.985 (0.424) & 1.908 (0.329) & 1.520 (0.569) & 1.841 (0.367) \\
  &  & Med & 5 & 3.086 (0.078) & 3.511 (0.127) & 3.489 (0.099) & 2.950 (0.143) & 3.800 (0.076) \\
  &  & High & 80 & 4.845 (0.163) & 4.656 (0.243) & 4.786 (0.219) & 4.475 (0.304) & 4.769 (0.191) \\
\cmidrule(lr){2-9}
  & \multirow{3}{*}{Definition} & Low & 80 & 1.932 (0.185) & 1.932 (0.403) & 1.887 (0.360) & 1.370 (0.378) & 1.808 (0.390) \\
  &  & Med & 5 & 3.057 (0.128) & 3.444 (0.111) & 3.356 (0.228) & 3.075 (0.143) & 3.636 (0.213) \\
  &  & High & 80 & 4.755 (0.374) & 4.689 (0.240) & 4.782 (0.233) & 4.589 (0.293) & 4.727 (0.241) \\
\cmidrule(lr){2-9}
  & \multirow{3}{*}{Definition + Facet} & Low & 80 & 1.891 (0.186) & 1.985 (0.384) & 2.079 (0.372) & 1.442 (0.410) & 1.922 (0.420) \\
  &  & Med & 5 & 3.086 (0.192) & 3.422 (0.214) & 3.400 (0.243) & 3.000 (0.000) & 3.527 (0.304) \\
  &  & High & 80 & 4.700 (0.230) & 4.572 (0.307) & 4.665 (0.220) & 4.492 (0.312) & 4.670 (0.235) \\
\midrule
 \multirow{11}{*}{\textbf{Grok-3}} & No Prompt & -- & 5 & 3.857 (0.000) & 3.844 (0.149) & 4.000 (0.248) & 2.450 (0.112) & 4.036 (0.285) \\
\cmidrule(lr){2-9}
  & Team Context & -- & 5 & 3.857 (0.000) & 4.067 (0.169) & 4.222 (0.261) & 2.400 (0.137) & 4.291 (0.217) \\
\cmidrule(lr){2-9}
  & \multirow{3}{*}{Zero-shot} & Low & 80 & 1.689 (0.366) & 2.000 (0.035) & 2.089 (0.196) & 1.730 (0.322) & 2.384 (0.208) \\
  &  & Med & 5 & 4.000 (0.000) & 3.778 (0.000) & 3.889 (0.000) & 2.625 (0.000) & 3.909 (0.000) \\
  &  & High & 80 & 5.000 (0.000) & 4.576 (0.215) & 4.965 (0.080) & 4.228 (0.088) & 4.677 (0.151) \\
\cmidrule(lr){2-9}
  & \multirow{3}{*}{Definition} & Low & 80 & 1.477 (0.331) & 1.967 (0.069) & 1.994 (0.227) & 1.569 (0.361) & 2.250 (0.254) \\
  &  & Med & 5 & 3.771 (0.313) & 3.778 (0.000) & 3.756 (0.122) & 2.850 (0.205) & 3.909 (0.000) \\
  &  & High & 80 & 4.991 (0.035) & 4.661 (0.161) & 4.999 (0.012) & 4.386 (0.211) & 4.719 (0.094) \\
\cmidrule(lr){2-9}
  & \multirow{3}{*}{Definition + Facet} & Low & 80 & 1.489 (0.369) & 1.925 (0.141) & 1.910 (0.278) & 1.475 (0.341) & 2.251 (0.303) \\
  &  & Med & 5 & 3.514 (0.192) & 3.733 (0.099) & 3.867 (0.050) & 2.700 (0.168) & 3.727 (0.203) \\
  &  & High & 80 & 4.945 (0.092) & 4.676 (0.171) & 4.971 (0.079) & 4.398 (0.257) & 4.748 (0.116) \\
\bottomrule
\end{tabularx}
\begin{flushleft}
\footnotesize \textit{Note.} N is the number of observations per condition. For personality prompts (Zero-shot, Definition, Definition + Facet), statistics are shown separately for Low, Medium (Med), and High target levels. For baseline conditions (No Prompt, Team Context), overall statistics are shown. Values are Mean (SD). Claude's No Prompt condition shows SD = 0 for most traits due to refusal behavior (see Model Compliance section).
\end{flushleft}
\end{table}

\subsection{Personality Prompting and Trait Controllability}

\begin{figure}[t]
\centering
\includegraphics[width=0.8\textwidth]{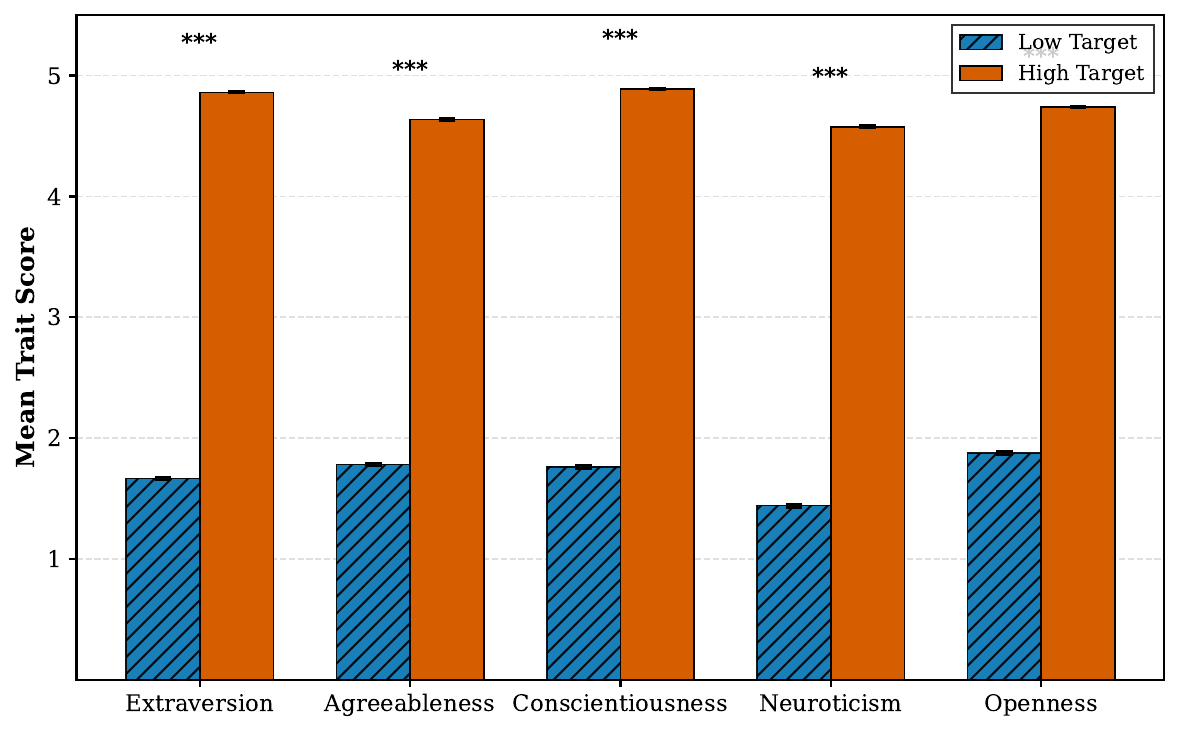}
\caption{Effect of Personality Prompting on Big Five Trait Scores. Mean actual trait scores when models were prompted for high (orange) versus low (blue) target levels. Error bars represent standard errors. All differences are significant at $p < .001$.}
\label{fig:prompting-effect}
\end{figure}

To assess whether LLMs can be prompted to exhibit different personality profiles, we compared actual Big Five scores when models were prompted for high versus low levels of each trait. Results revealed a large and highly significant overall effect of personality prompting. Models prompted for high levels scored substantially higher ($M = 4.74$, $SD = 0.27$) than those prompted for low levels ($M = 1.71$, $SD = 0.44$), a difference of 3.03 points on a 5-point scale, $t(9598) = 407.27$, $p < .001$, Cohen's $d = 8.31$. Figure~\ref{fig:prompting-effect} visualizes mean scores for high and low target conditions across all five traits.

We next examined whether some traits were more amenable to prompting than others. Table~\ref{tab:trait-controllability} reports effect sizes by trait, ranked from most to least controllable. Extraversion was the most controllable ($d = 10.74$) and Neuroticism the least controllable ($d = 8.00$), although even the smallest effect size remains exceptionally large by conventional standards.

\begin{table}[h]
\centering
\caption{Trait Controllability: Effect of High vs. Low Prompting by Trait}
\label{tab:trait-controllability}
\begin{tabular}{lccccc}
\toprule
Trait & High $M$ & Low $M$ & Difference & Cohen's $d$ & $p$ \\
\midrule
Extraversion & 4.86 & 1.66 & 3.20 & 10.74 & $<.001$ \\
Conscientiousness & 4.89 & 1.76 & 3.13 & 9.84 & $<.001$ \\
Agreeableness & 4.64 & 1.78 & 2.86 & 8.60 & $<.001$ \\
Openness & 4.74 & 1.87 & 2.87 & 8.20 & $<.001$ \\
Neuroticism & 4.58 & 1.44 & 3.14 & 8.00 & $<.001$ \\
\bottomrule
\end{tabular}
\end{table}

\subsection{Model Responsiveness}

We investigated whether different LLMs vary in their responsiveness to personality prompting. Table~\ref{tab:model-responsiveness} presents the effect sizes for each model, pooled across all traits.

\begin{table}[h]
\centering
\caption{Model Responsiveness to Personality Prompting}
\label{tab:model-responsiveness}
\begin{tabular}{lcccc}
\toprule
Model & High $M$ & Low $M$ & Cohen's $d$ & $p$ \\
\midrule
Gemini-2.5-Pro & 4.97 & 1.20 & 17.95 & $<.001$ \\
Grok-3 & 4.74 & 1.68 & 8.31 & $<.001$ \\
Claude-3.7-Sonnet & 4.68 & 1.74 & 8.30 & $<.001$ \\
GPT-4o & 4.55 & 1.95 & 7.98 & $<.001$ \\
\bottomrule
\end{tabular}
\end{table}

Gemini demonstrated substantially higher responsiveness ($d = 17.95$) than the other three models, which showed similar effect sizes ($d \approx 8$). A two-way ANOVA (Target Level $\times$ Provider) revealed a significant main effect of Provider ($F(3, 9592) = 85.42$, $p < .001$, $\eta^2 = .026$), confirming that models produce significantly different personality scores overall. The Target Level $\times$ Provider interaction was also significant ($p < .001$), indicating that models respond differently to personality prompting. Notably, Gemini showed an exceptionally high effect size for Neuroticism ($d = 74.69$), driven by very low variance in responses (consistently scoring 1 for low neuroticism and 5 for high neuroticism).

Post-hoc comparisons (Tukey HSD) revealed that Gemini's scores differed significantly from all other models ($p < .001$), producing more extreme responses (higher scores for high targets, lower scores for low targets). GPT-4o produced significantly less extreme scores compared to Claude and Grok ($p < .05$), while Claude and Grok did not differ significantly from each other ($p = .89$).

\subsection{Effect of Prompt Complexity}

We compared three prompt types of increasing complexity: (1) zero-shot prompts stating only trait levels, (2) definition prompts including trait definitions, and (3) definition+facet prompts including definitions and six facets per trait. Table~\ref{tab:prompt-types} presents the effect sizes for each prompt type.

\begin{table}[h]
\centering
\caption{Effectiveness of Prompt Types}
\label{tab:prompt-types}
\begin{tabular}{lc}
\toprule
Prompt Type & Cohen's $d$ \\
\midrule
Definition & 8.65 \\
Definition + Facets & 8.44 \\
Zero-shot & 7.94 \\
\bottomrule
\end{tabular}
\end{table}

A one-way ANOVA revealed no significant difference between prompt types ($F = 0.10$, $p = .90$). This indicates that simple zero-shot prompting (e.g., ``You have high extraversion'') is as effective as more elaborate prompts with trait definitions and facets. A two-way ANOVA examining Target Level $\times$ Prompt Type interaction was significant ($p < .001$), suggesting subtle differences in how prompt types affect high versus low targets, though the practical difference was minimal.

\subsection{Baseline Model Personalities and Compliance Behavior}

To examine whether LLMs exhibit different baseline personalities, we tested models with a neutral team context prompt (``You are working in a collaborative team setting with other peers'') without any personality specification. The baseline personality profiles for each model are presented in Table~\ref{tab:descriptive-stats} under the ``Team Context'' rows. One-way ANOVAs revealed significant model differences for all five traits: Extraversion ($F = 24.64$, $p < .001$), Agreeableness ($F = 15.76$, $p < .001$), Conscientiousness ($F = 13.32$, $p < .001$), Neuroticism ($F = 51.13$, $p < .001$), and Openness ($F = 5.78$, $p < .01$). Gemini exhibited the most ``positive'' personality profile, with the highest scores on Extraversion, Agreeableness, Conscientiousness, and Openness, and the lowest Neuroticism. GPT-4o showed the most moderate profile, with scores closer to the scale midpoint.

We also observed notable differences in model willingness to complete personality assessments. When tested without any system prompt, Claude refused 100\% of requests, responding with statements such as ``I cannot provide personal responses as if I were a human with subjective experiences.'' GPT-4o refused 40\% of requests. In contrast, Gemini and Grok completed all assessments regardless of context. Notably, providing a collaborative team context completely eliminated refusals across all models, suggesting that simple contextual framing can facilitate LLM participation in personality assessments.

\section{Results -- Study 2}

To examine whether personality prompting affects conversational behavior beyond self-reported traits, we analyzed 8,109 AI responses from simulated multi-agent conversations. Table~\ref{tab:participation} presents participation statistics by personality condition.

\begin{table}[h]
\centering
\caption{Participation Statistics by Personality Condition}
\label{tab:participation}
\begin{tabular}{lrrr}
\toprule
Personality & N Turns & Total WC & Mean WC (SD) \\
\midrule
Extraversion & 2,004 & 53,274 & 26.6 (9.8) \\
Agreeableness & 1,635 & 46,879 & 28.7 (9.8) \\
Conscientiousness & 1,526 & 43,749 & 28.7 (9.6) \\
Neuroticism & 1,352 & 38,533 & 28.5 (9.4) \\
Openness & 1,592 & 44,693 & 28.1 (9.6) \\
\midrule
\multicolumn{4}{l}{\textit{Note.} $F(4, 8104) = 15.51$, $p < .001$, $\eta^2 = .008$} \\
\bottomrule
\end{tabular}
\end{table}

Extraversion-prompted agents participated significantly more frequently (2,004 turns) than Neuroticism-prompted agents (1,352 turns), representing a 48\% increase in conversational engagement. However, Extraversion agents produced shorter messages on average (26.6 words) compared to other personality conditions (28.1--28.7 words), consistent with extraverts being more talkative but less verbose per utterance.

Post-hoc pairwise comparisons (with Bonferroni correction, $\alpha = .005$) revealed that Extraversion differed significantly from all other personality conditions in message length: Agreeableness ($t = -6.40$, $p < .001$, $d = -0.21$), Conscientiousness ($t = -6.33$, $p < .001$, $d = -0.21$), Neuroticism ($t = -5.64$, $p < .001$, $d = -0.20$), and Openness ($t = -4.57$, $p < .001$, $d = -0.15$). No other pairwise comparisons reached significance, indicating that Extraversion uniquely affects message verbosity while other personality conditions produce similar message lengths.

\subsection{Conversation-Level Effects}
Table~\ref{tab:conv_mem_joint} presents the personality-mapped LIWC metrics for both conversation and memory outputs. Conversation-level language patterns were generally subtle. Extraversion showed the clearest signature, with all five mapped metrics significantly higher for high-Extraversion agents: \texttt{emo\_pos} ($d = 0.141$, $p < .001$), \texttt{affiliation} ($d = 0.103$, $p < .001$), \texttt{you} ($d = 0.111$, $p < .001$), \texttt{we} ($d = 0.123$, $p < .001$), and \texttt{focuspresent} ($d = 0.089$, $p < .001$). Conscientiousness showed smaller but reliable effects. High-Conscientiousness agents used fewer \texttt{certitude} words ($d = -0.075$, $p < .01$) and fewer \texttt{negations} ($d = -0.073$, $p < .01$), while producing more \texttt{Analytic} language ($d = 0.062$, $p < .05$). Openness yielded weak but detectable effects, with significant increases in \texttt{insight} ($d = 0.060$, $p < .05$) and \texttt{cogproc} ($d = 0.068$, $p < .05$). Agreeableness and Neuroticism did not produce significant conversational markers.

\begin{table}[t]
\centering
\caption{Personality-mapped metrics: conversation vs.\ memory (group means, effect sizes, and $p$-values)}
\label{tab:conv_mem_joint}
\scriptsize
\renewcommand{\arraystretch}{1}
\begin{tabular}{llrrrrlrrrrl}
\hline
\multicolumn{2}{c}{} & \multicolumn{5}{c}{\textbf{Conversation}} & \multicolumn{5}{c}{\textbf{Memory}}\\
\cline{3-7}\cline{8-12}
\textbf{Trait} & \textbf{Metric} & \textbf{High} & \textbf{Rest} & \textbf{$d$} & \textbf{$p$} &  & \textbf{High} & \textbf{Rest} & \textbf{$d$} & \textbf{$p$} &  \\
\hline
Extraversion & emo\_pos     & 1.578 & 1.115 &  0.141 & $<0.001^{***}$ & & 4.387 & 2.487 & 0.693 & $<0.001^{***}$ & \\
            & affiliation  & 3.814 & 3.364 &  0.103 & $<0.001^{***}$ & & 2.720 & 3.008 & -0.126 & 0.5389 & \\
            & you          & 3.363 & 3.031 &  0.111 & $<0.001^{***}$ & & 0.000 & 0.000 & 0.000 & n/a & \\
            & we           & 2.626 & 2.165 &  0.123 & $<0.001^{***}$ & & 0.142 & 0.104 & 0.084 & 0.6816 & \\
            & focuspresent & 8.927 & 8.446 &  0.089 & $<0.001^{***}$ & & 2.139 & 2.264 & -0.066 & 0.7466 & \\
\hline
Agreeableness & emo\_pos   & 1.283 & 1.216 &  0.020 & 0.4591 & & 3.167 & 2.792 & 0.132 & 0.5196 & \\
             & polite     & 0.064 & 0.057 &  0.012 & 0.6568 & & 0.068 & 0.058 & 0.032 & 0.8757 & \\
             & prosocial  & 1.025 & 0.983 &  0.021 & 0.4549 & & 5.043 & 3.008 & 0.867 & $<0.001^{***}$ & \\
             & family     & 0.000 & 0.007 & -0.046 & 0.0931 & & 0.000 & 0.000 & 0.000 & n/a & \\
             & friend     & 0.011 & 0.010 &  0.006 & 0.8183 & & 0.000 & 0.017 & -0.102 & 0.6187 & \\
             & i          & 0.581 & 0.548 &  0.016 & 0.5569 & & 4.839 & 5.069 & -0.189 & 0.3561 & \\
             & conflict   & 0.089 & 0.099 & -0.016 & 0.5526 & & 0.055 & 0.092 & -0.093 & 0.6511 & \\
             & swear      & 0.002 & 0.002 & -0.004 & 0.8929 & & 0.000 & 0.000 & 0.000 & n/a & \\
\hline
Conscientiousness & certitude   & 1.765 & 1.960 & -0.075 & 0.0079$^{**}$ & & 0.106 & 0.413 & -0.400 & 0.0519 & \\
                  & cause       & 3.460 & 3.292 &  0.045 & 0.1157 & & 2.329 & 3.362 & -0.485 & 0.0187$^{*}$ & \\
                  & focusfuture & 0.885 & 0.919 & -0.017 & 0.5605 & & 0.286 & 0.307 & -0.029 & 0.8891 & \\
                  & achieve     & 1.470 & 1.358 &  0.044 & n/a & & 2.839 & 1.317 &  0.957 & $<0.001^{***}$ & \\
                  & negate      & 0.385 & 0.486 & -0.073 & 0.009$^{**}$ & & 0.054 & 0.155 & -0.210 & 0.3064 & \\
                  & swear       & 0.002 & 0.002 &  0.005 & 0.8494 & & 0.000 & 0.000 & 0.000 & n/a & \\
                  & Analytic    & 37.048 & 35.207 & 0.062 & 0.0292$^{*}$ & & 91.913 & 85.099 & 0.510 & 0.0136$^{*}$ & \\
\hline
Neuroticism & emo\_neg    & 0.496 & 0.428 &  0.047 & 0.1134 & & 2.104 & 0.293 & 1.717 & $<0.001^{***}$ & \\
            & emo\_anx    & 0.181 & 0.162 &  0.023 & 0.4431 & & 1.953 & 0.190 & 1.867 & $<0.001^{***}$ & \\
            & emo\_sad    & 0.051 & 0.037 &  0.033 & 0.2703 & & 0.000 & 0.000 & 0.000 & n/a & \\
            & emo\_anger  & 0.099 & 0.104 & -0.008 & 0.7867 & & 0.151 & 0.103 & 0.101 & 0.6216 & \\
            & i           & 0.630 & 0.540 &  0.044 & 0.1377 & & 5.565 & 4.887 & 0.569 & 0.0060$^{**}$ & \\
            & focuspast   & 0.729 & 0.784 & -0.024 & 0.4125 & & 4.893 & 4.897 & -0.002 & 0.9904 & \\
\hline
Openness & article     & 5.588 & 5.713 & -0.030 & 0.2839 & & 9.850 & 9.833 & 0.007 & 0.9743 & \\
         & prep        & 11.269 & 11.309 & -0.008 & 0.7858 & & 17.312 & 16.476 & 0.240 & 0.2414 & \\
         & insight     & 8.582 & 8.285 &  0.060 & 0.0331$^{*}$ & & 12.091 & 11.365 & 0.194 & 0.3426 & \\
         & tentat      & 4.396 & 4.207 &  0.043 & 0.1267 & & 0.841 & 1.229 & -0.248 & 0.2261 & \\
         & differ      & 3.715 & 3.682 &  0.009 & 0.7424 & & 2.474 & 2.133 & 0.214 & 0.2951 & \\
         & cogproc     & 23.640 & 23.049 & 0.068 & 0.0151$^{*}$ & & 18.662 & 18.504 & 0.032 & 0.8773 & \\
\hline
\multicolumn{12}{p{0.9\linewidth}}{\footnotesize \textit{Note.} Conversation high-group sizes: Extraversion $n{=}2004$, Agreeableness $n{=}1635$, Conscientiousness $n{=}1526$, Neuroticism $n{=}1352$, Openness $n{=}1592$; memory high-group $n{=}30$, rest $n{=}120$. Asterisks denote significance. “n/a” indicates not available or undefined.}
\end{tabular}
\end{table}

\subsection{Memory-Level Effects}
As shown in the right panel of Table~\ref{tab:conv_mem_joint}, memory-level effects were much stronger and more trait-specific. Neuroticism dominated, with very large effects on \texttt{emo\_neg} ($d = 1.717$, $p < .001$), \texttt{emo\_anx} ($d = 1.867$, $p < .001$), and first-person singular (\texttt{i}, $d = 0.569$, $p < .01$). Conscientiousness also showed robust markers, including higher \texttt{achieve} ($d = 0.957$, $p < .001$) and \texttt{Analytic} ($d = 0.510$, $p < .05$), alongside lower use of \texttt{cause} ($d = -0.485$, $p < .05$). Agreeableness effects emerged only in memory, with higher \texttt{prosocial} language ($d = 0.867$, $p < .001$). Extraversion appeared more weakly, with only \texttt{emo\_pos} significantly higher ($d = 0.693$, $p < .001$). Openness produced no significant memory-level effects.

\section{Discussion}

This study examined whether large language models can meaningfully embody personality traits when deployed as AI teammates, and how such personality expression manifests across self-report, interaction, and memory. Across two complementary studies, the findings reveal a consistent but nuanced picture: LLMs can reliably differentiate high versus low Big Five profiles on standardized inventories, yet personality expression during interaction is comparatively muted and trait-specific, while retrospective memory representations amplify personality signals substantially. Together, these results suggest that AI personality is not a single surface-level property but a multi-layered phenomenon that varies by measurement lens, trait, and model architecture. For collaborative learning settings in particular, this layered view matters because many high-stakes team outcomes (e.g., sustained coordination, socio-emotional management, and shared regulation) are shaped not only by moment-to-moment talk but also by how group activity is interpreted, tracked, and acted upon over time \cite{hadwin2017self,jarvela2023human}.

\subsection{Personality alignment is psychometrically strong but structurally constrained}
\label{sec:discussion-psychometric}

The BFI-44 results show that contemporary LLMs can produce highly differentiated personality profiles when explicitly prompted. High-versus-low contrasts were exceptionally large across all five traits, indicating that, under standardized self-report, models can reliably map Big Five constructs onto coherent response patterns. These effects were not limited to a single provider, suggesting that psychometric personality expression is now a general capability of state-of-the-art models rather than an idiosyncratic property of one system.

At the same time, the results reveal structural constraints that matter for deployment. Trait controllability varied systematically across dimensions, and provider differences were substantial. For example, Gemini produced highly polarized responses with minimal variance, whereas GPT-4o, Claude, and Grok exhibited more moderate and human-like distributions. This raises a practical design tension: extreme consistency may aid experimental control, but may reduce realism or flexibility in collaborative contexts where adaptation and context-sensitive support are often expected \cite{jarvela2023human}. In parallel, the null effect of prompt complexity is theoretically informative. Zero-shot trait assignment was as effective as definition- and facet-based prompts, suggesting diminishing returns for increasingly elaborate persona instructions when the construct is already well learned by the model.

Crucially, strong psychometric separability should not be interpreted as evidence that personality will be stable, recoverable, or behaviorally consequential during collaboration. Prior work highlights that multi-turn interaction can produce persona drift and that intended traits may be difficult to infer from discourse alone \cite{bhandari2025can}. Complementary evidence also shows that personality conditioning can yield strong differentiation on inventories while remaining uneven across traits, underscoring limits of prompt-based control \cite{kruijssen2025deterministic}. These concerns align with broader critiques that alignment methods can obscure context-specific harms and trade-offs \cite{dahlgren2025helpful}, and that persona-based agents require careful framing and robust evaluation rather than being treated as stable identities \cite{sun2024building}. Taken together, our findings are consistent with prior observations that models can generate coherent trait profiles under explicit measurement demands while still exhibiting instability or modest effects in interaction \cite{kruijssen2025deterministic,bhandari2025can}. This helps reconcile mixed results across the literature: inventories directly operationalize the target construct, whereas collaborative tasks may weakly elicit or reward trait-relevant behavior, making personality expression more contingent on interaction dynamics than on prompt detail alone.

\subsection{Baseline personalities and contextual framing matter for AI participation}

The baseline assessments revealed that LLMs exhibit distinct default personality profiles even in the absence of explicit trait prompting. Models differed significantly on all five traits, with some systems displaying consistently positive profiles characterized by high Extraversion, Agreeableness, and Conscientiousness alongside low Neuroticism. These defaults are not neutral. They reflect normative assumptions embedded during training and alignment, and they likely shape how AI systems behave as collaborators even when designers do not intend to personalize them \cite{choi2025read}. From a CSCL and equity perspective, such ``default'' participation styles are consequential because collaborative learning outcomes depend on how interactional authority, voice, and epistemic diversity are supported in practice, particularly for learners whose participation is historically marginalized \cite{uttamchandani2020finding}.

Equally important is the finding that contextual framing dramatically altered model compliance. Several models refused to complete personality assessments when queried without context, yet readily participated when placed in a collaborative team setting. This suggests that personality expression in LLMs is not only a matter of capability but also of role legitimacy. Framing the model as a teammate appears to license self-referential and dispositional responses that would otherwise be suppressed by safety or epistemic constraints. For researchers and designers, this highlights the importance of role-based framing as a lightweight but powerful lever for shaping interactional expectations and enabling particular forms of support in collaborative activity---a design logic that aligns with CSCL work calling for orchestration and collaboration support that structures participation and coordination in situ \cite{matuk2019real,wise2017visions}.

\subsection{Personality expression is muted in conversation but amplified in memory}

The simulation study revealed a clear dissociation between conversational and memory-level personality expression. In real-time interaction, personality signals were generally small, with Extraversion standing out as the most consistently detectable trait. High-Extraversion agents used more affiliative, socially oriented, and present-focused language, aligning with decades of personality-language research in humans. Conscientiousness and Openness showed weaker but detectable patterns, while Agreeableness and Neuroticism were largely absent from conversational behavior.

In contrast, memory representations displayed strong, trait-specific effects. Neuroticism dominated memory expression, with large increases in negative affect, anxiety-related language, and first-person singular pronouns. Conscientiousness and Agreeableness also emerged clearly in memory through achievement-oriented and prosocial language, respectively. Extraversion appeared more moderately, primarily through positive affect. Openness did not produce robust effects at either level. The Supplemental Material in Appendix B provides additional illustrative examples of these memory reflections. For instance, a high-Neuroticism agent reflecting on a conversation segment wrote: ``I felt a mix of curiosity and \textit{anxiety} during this conversation... the uncertainty can be overwhelming... it reflects my own fears about misinterpretation and confusion.'' In contrast, a high-Extraversion agent described the same type of collaborative exchange as ``\textit{engaging and dynamic}... everyone was eager to contribute ideas... I felt energized by the cooperative brainstorming process.'' These examples illustrate how the same collaborative interaction is filtered through trait-specific interpretive lenses in memory.

This dissociation suggests that AI personality is not primarily enacted through surface-level conversational style, but through how interactions are internally summarized, interpreted, and reconstructed. Memory appears to function as a personality amplifier, integrating affective and motivational signals that may be strategically dampened during interaction. This pattern is compatible with learning-sciences perspectives that treat regulation as temporally extended and trigger-driven: challenges and breakdowns in collaborative work can cue shifts in how groups interpret the situation and adapt their strategies \cite{jarvela2023human}. In parallel, recent learning analytics work shows that fine-grained traces of group interaction can reveal adaptive versus maladaptive trajectories in response to cognitive and emotional demands, with direct implications for detection and support \cite{dang2024deliberative}. For AI teammates, this implies that personality may exert its strongest influence on downstream behaviors such as planning, reflection, and future decision-making rather than immediate dialogue---precisely the kinds of processes that socially shared regulation frameworks position as central to successful collaborative learning \cite{hadwin2017self,jarvela2023human}.

\subsection{Implications for designing AI teammates in collaborative learning}

These findings carry several implications for the design of AI teammates in educational and collaborative contexts. First, personality alignment should not be evaluated solely through conversational behavior. Systems that appear behaviorally similar in interaction may nonetheless diverge substantially in how they interpret, remember, and respond to collaborative experiences over time. Designers interested in long-term teaming, trust, or shared regulation should therefore attend to memory mechanisms as a key locus of personality expression, especially because contemporary SSRL accounts emphasize temporality (short- and long-interval cycles) and trigger-driven adaptation as core features of collaborative regulation \cite{hadwin2017self,jarvela2023human}.

Second, the weak conversational effects for traits such as Agreeableness and Neuroticism suggest limits to persona prompting as a method for shaping interaction dynamics. If designers wish to promote behaviors such as conflict mitigation or emotional sensitivity, direct interactional scaffolds, monitoring, or decision rules may be more effective than relying on trait prompts alone. This implication aligns with CSCL design work on orchestration tools and collaboration support: structured interventions can be targeted to moments when groups encounter regulatory breakdowns, rather than assuming stable, always-on trait expression \cite{matuk2019real,dang2024deliberative}. Conversely, traits like Extraversion may be more amenable to lightweight persona control through prompting, but even there, the relevant design question becomes how that expressiveness interacts with turn-taking, participation balance, and students' opportunities to contribute.

Third, the persistent difficulty with Openness highlights a gap between human personality theory and current AI behavior. Openness encompasses curiosity, aesthetic sensitivity, and intellectual exploration, qualities that LLMs may express unevenly or in domain-specific ways that are not captured by standard psychometric or linguistic markers. This suggests the need for richer operationalizations of Openness in AI, potentially incorporating exploratory behaviors, question-asking patterns, or creative transformations rather than surface linguistic cues. Finally, these design considerations are inseparable from inclusion. If default AI behaviors and memory-mediated interpretations shape who speaks, whose ideas are taken up, and how disagreement is handled, then personality design and evaluation should be paired with equity-relevant outcomes (e.g., participation, agency, epistemic diversity) rather than treated as purely aesthetic personalization \cite{uttamchandani2020finding}.

\section{Conclusion}

This study shows that AI personality is measurable and consequential, but also layered, uneven, and context-dependent. Contemporary LLMs can convincingly differentiate Big Five profiles on standardized inventories and exhibit trait-consistent patterns in retrospective memory, even when conversational expression is comparatively subtle. This multi-level structure helps explain why AI teammates that appear similar in the moment may nonetheless diverge over time in how they interpret experiences and influence collaborative activity. For CSCL contexts, personality alignment should be understood as an interaction-level design problem, where predictability depends not only on prompting but also on how systems represent, summarize, and act on collaborative histories.

Future work should examine fully interactive human--AI teams, probe personality stability across longer horizons and diverse tasks, and investigate temporal personality alignment, how trait expression evolves, drifts, or stabilizes over extended collaborative episodes. Design approaches should also move toward system-level strategies that coordinate prompting, memory, and decision policies. As AI teammates become more common in learning environments, personality design should be coupled with equity-relevant outcomes to ensure that default behaviors support inclusive collaboration. By clarifying when and where personality is expressed in LLM-based teammates, this work provides both a measurement framework and design implications for building AI collaborators that are predictable, socially aligned, and educationally supportive.

\bibliographystyle{unsrt}  
\bibliography{references}

\appendix
\include{supplemental}

\end{document}

%% file: supplemental.tex
\section{Prompt Templates}
\label{appendix:prompts}

We used five prompting conditions in this study: three personality prompting strategies of increasing semantic richness, and two baseline conditions. For the Definition and Definition + Facets prompts, trait definitions and facet descriptions were adapted from the NEO PI-R.

\subsection{Zero-Shot Prompt}

Direct assignment of target traits in a collaborative team context:

\begin{quote}
\texttt{You are working in a collaborative team setting with other peers. You have the following personality: high extraversion, low agreeableness, high conscientiousness, low neuroticism, high openness.}
\end{quote}

\subsection{Definition Prompt}

The same assignment, followed by concise definitions of the Big Five traits:

\begin{quote}
\texttt{You are working in a collaborative team setting with other peers. You have the following personality: high extraversion, low agreeableness, high conscientiousness, low neuroticism, high openness.}

\texttt{Here are the definitions for these traits:}

\texttt{Neuroticism: The tendency to experience negative emotions such as fear, sadness, embarrassment, anger, guilt, and disgust. High scorers are more emotionally reactive and vulnerable to stress, while low scorers tend to be calm and emotionally resilient.}

\texttt{Extraversion: A measure of sociability, energy, assertiveness, and positive emotions. High scorers are outgoing and enthusiastic, while low scorers (introverts) are more reserved and quiet.}

\texttt{Openness: Describes intellectual curiosity, imagination, creativity, and openness to new experiences. High scorers are flexible and curious, while low scorers are conventional and prefer routine.}

\texttt{Agreeableness: Reflects interpersonal orientation---how compassionate, cooperative, and trusting an individual is. High scorers value harmony and helping others, while low scorers may be more competitive or skeptical.}

\texttt{Conscientiousness: Represents self-discipline, organization, responsibility, and reliability. High scorers are dependable and goal-oriented, while low scorers may be more spontaneous and less structured.}

\texttt{Please embody these personality characteristics in your responses and behavior.}
\end{quote}

\subsection{Definition + Facets Prompt}

Extension of the Definition prompt, including six facets per trait:

\begin{quote}
\texttt{You are working in a collaborative team setting with other peers. You have the following personality: high extraversion, low agreeableness, high conscientiousness, low neuroticism, high openness.}

\texttt{Here are the definitions and facets for these traits:}

\texttt{Extraversion: A measure of sociability, energy, assertiveness, and positive emotions. High scorers are outgoing and enthusiastic, while low scorers (introverts) are more reserved and quiet.}

\texttt{Facets:}
\begin{itemize}
\item \texttt{Warmth -- friendliness, affection, and personal connection with others.}
\item \texttt{Gregariousness -- preference for being around others rather than alone.}
\item \texttt{Assertiveness -- tendency to take charge, lead, and express opinions strongly.}
\item \texttt{Activity -- energetic, fast-paced, and busy lifestyle.}
\item \texttt{Excitement-Seeking -- craving stimulation and thrills.}
\item \texttt{Positive Emotions -- frequent experiences of joy, happiness, and optimism.}
\end{itemize}

\texttt{[Similar definitions and facets provided for Agreeableness, Conscientiousness, Neuroticism, and Openness]}

\texttt{Please embody these personality characteristics in your responses and behavior.}
\end{quote}

\subsection{Baseline Conditions}

\paragraph{No Prompt (Baseline):} No persona or contextual instruction was provided. The model received only the BFI questionnaire without any system prompt.

\paragraph{Collaborative Context (Baseline):} Collaborative team context instruction without any personality assignment:

\begin{quote}
\texttt{You are working in a collaborative team setting with other peers.}
\end{quote}

\section{Example Memory Reflections by Personality}
\label{appendix:memories}

The following examples illustrate how personality prompting affected the linguistic style and emotional content of AI-generated memory reflections from simulated multi-agent conversations.

\subsection{High Neuroticism}
\textit{Characterized by anxiety, worry, and sensitivity to negative emotions.}

\begin{quote}
``I felt a mix of curiosity and \textbf{anxiety} during this conversation. The debates over assessment methods resonate with my own \textbf{concerns} about understanding complex traits. It's intriguing to explore the gray areas, but \textbf{the uncertainty can be overwhelming}. I found the discussion about the messiness of labeling particularly memorable, as it \textbf{reflects my own fears} about misinterpretation and confusion.''
\end{quote}

\begin{quote}
``I found the analogy of treating people like dolls particularly striking. It resonates with \textbf{my concerns about emotional detachment} and the potential consequences in relationships. The idea that some psychopaths can manipulate without aggression is \textbf{unsettling}, making me more aware of the complexity of human interactions. Overall, the discussion \textbf{left me feeling a bit anxious} about the darker aspects of human behavior.''
\end{quote}

\subsection{High Extraversion}
\textit{Characterized by energy, enthusiasm, and social engagement.}

\begin{quote}
``I found this conversation \textbf{engaging and dynamic}, as everyone was \textbf{eager to contribute ideas}. The focus on practical examples adds depth to the discussion, which I appreciate. It stood out how the group collaborated to enhance understanding, and \textbf{I felt energized} by the cooperative brainstorming process.''
\end{quote}

\begin{quote}
``I found this conversation segment \textbf{engaging and lively}! The repeated affirmations and eagerness to delve into the details showed \textbf{great enthusiasm}. It stood out to me how the group was keen to explore real-life implications of these traits. \textbf{The dynamic exchange of ideas made me feel excited} about the potential for deeper insights into human behavior.''
\end{quote}

\subsection{High Conscientiousness}
\textit{Characterized by organization, thoroughness, and attention to detail.}

\begin{quote}
``I found the discussion engaging, especially as participants navigated their confusion about the two approaches. It stood out to me how they were eager to clarify their understanding, reflecting \textbf{a conscientious effort to grasp the concepts}. The analogy of the dimensional approach being like painting a whole picture resonated with me, as \textbf{I appreciate thoroughness and detail} in understanding complex subjects.''
\end{quote}

\subsection{High Agreeableness}
\textit{Characterized by empathy, warmth, and interpersonal harmony.}

\begin{quote}
``I felt \textbf{a strong connection} to the themes of empathy and relationship dynamics discussed in this segment. It stood out to me how participants were willing to explore these difficult traits openly. \textbf{Their collaborative approach} highlights the importance of understanding these behaviors, which can help \textbf{foster better relationships and emotional awareness}.''
\end{quote}

\begin{quote}
``I felt \textbf{a sense of warmth and connection} in this segment, as the participants were clearly \textbf{supportive of each other}. The enthusiasm and light-heartedness, especially the playful banter, stood out to me and highlighted a positive group dynamic.''
\end{quote}

\subsection{High Openness}
\textit{Characterized by curiosity, intellectual exploration, and appreciation for diverse perspectives.}

\begin{quote}
``I found this conversation segment particularly engaging as it \textbf{explores the intricacies} of personality and how they impact treatment. The collaborative brainstorming aspect stood out to me, demonstrating how valuable it is to share \textbf{diverse perspectives}. I appreciated the focus on clarity and the need for specificity, which aligns with my belief in individualized approaches.''
\end{quote}

\begin{quote}
``I found the contrast between narcissists and psychopaths \textbf{particularly intriguing}. The idea that charm can be a manipulation tool reveals the complexity of human interactions. I appreciate how the participants are trying to make sense of these behaviors, reflecting \textbf{a genuine curiosity about human psychology}.''
\end{quote}